\documentstyle[aps,pre,epsf,floats]{revtex}
\begin{document}
\def\theequation{\arabic{section}.\arabic{equation}}
\global\firstfigfalse

\newcommand{\beq}{\begin{equation}}
\newcommand{\eeq}{\end{equation}}
\newcommand{\eqna}{\begin{eqnarray}}
\newcommand{\eqne}{\end{eqnarray}}
\newcommand{\dia}{\begin{displaymath}}
\newcommand{\die}{\end{displaymath}}
\newcommand{\eqnaa}{\begin{eqnarray*}}
\newcommand{\eqnae}{\end{eqnarray*}}

\def\dleft{\rlap{{\it D}}\raise 8pt\hbox{$\scriptscriptstyle\Leftarrow$}}
\def\prop{\propto}
\def\dright{\rlap{{\itD}}
\raise 8pt\hbox{$\scriptscriptstyle\Rightarrow$}}
\def\lrartop#1{#1\llap{
\raise 8pt\hbox{$\scriptscriptstyle\leftrightarrow$}}}
\def\_#1{_{\scriptscriptstyle #1}}
\def\&#1{^{\scriptscriptstyle #1}}
\def\sss{\scriptscriptstyle}
\def\dij{\delta_{\sss ij}}
\def\gij{g_{\sss ij}}
\def\Gij{g^{\sss ij}}
\def\gkm{g_{\sss km}}
\def\Gkm{g^{\sss km}}
\def\cd#1{{}_{\sss;#1}}
\def\ud#1{{}_{\sss,#1}}
\def\upcd#1{{}_{\sss;}{}^{\sss #1}}
\def\upud#1{{}_{\sss,}{}^{\sss #1}}
\def\ro{r\_{0}}
\def\vro{{\bf r}\_{0}}
\def\qo{q\_{1}}
\def\qt{q\_{2}}
\def\lfd{L\&{D}_f}
\def\rar{\rightarrow}
\def\deriv#1#2{{d#1\over d#2}}
\def\oot{{1\over 2}}
\def\pdline#1#2{\partial#1/\partial#2}
\def\pdd#1#2#3{{\partial#1\over\partial#2\partial#3}}
\def\av#1{\langle#1\rangle}
\def\avlar#1{\big\langle#1\big\rangle}
\def\div{{\vec\nabla}\cdot}
\def\grad{{\vec\nabla}}
\def\curl{{\vec\nabla}\times}
\def\A{{\cal A}}
\def\DD{{\cal D}}
\def\FF{{\cal F}}
\def\LL{{\cal L}}
\def\MM{{\cal M}}
\def\M{\lrartop{\MM}}
\def\PPs{{\cal P}}
\def\VV{{\cal V}}
\def\vd{\VV\_{D}}
\def\GG{{\cal G}}
\def\GGq{\GG_q}
\def\PP{\lrartop{\PPs}}
\def\EPS{{\cal E}}
\def\gf{\grad\f}
\def\rgf{\vec r\cdot\grad\f}
\def\bs{{\bf s}}
\def\bd{{\bf d}}
\def\be{{\bf e}}
\def\bff{{\bf f}}
\def\bF{{\bf F}}
\def\br{{\bf r}}
\def\bR{{\bf R}}
\def\bJ{{\bf J}}
\def\bV{{\bf V}}
\def\ba{{\bf a}}
\def\bA{{\bf A}}
\def\bg{{\bf g}}
\def\bn{{\bf n}}
\def\vh{{\bf h}}
\def\bI{{\bf I}}
\def\vI{\bI}
\def\gfds{\grad\f\cdot \bd\bs}
\def\hm{\hat\mu}
\def\hFF{\hat\FF}
\def\eegg{\be\otimes\be\grad\grad}
\def\inv{\int\_{V}}
\def\invv{\int\_{v}}
\def\ins{\int\_{\Sigma}}
\def\inss{\int\_{\sigma}}
\def\vi{v_i}
\def\sf{S\&{D}_f}
\def\si{S\&{D}_i}
\def\sd{S\&{D}}
\def\ds{\bd\bs}
\def\dtr{d\&3r}
\def\dDr{~d\&{D}r}
\def\dDh{~d\&{D}h}
\def\dtv{d\&3v}
\def\f{\varphi}
\def\fl{\varphi\_{L}}
\def\gfl{\grad\fl}
\def\fx{\f\_{x}}
\def\b{\beta}
\def\z{\zeta}
\def\bo{\beta\_{0}}
\def\bx{\beta\_{x}}
\def\d{\delta}
\def\c{\gamma}
\def\t{\tau}
\def\s{\sigma}
\def\r{\rho}
\def\l{\lambda}
\def\m{\mu}
\def\n{\nu}
\def\a{\alpha}
\def\abs#1{\vert #1\vert}
\def\abgf{\abs{\gf}}
\def\abgfl{\abs{\gfl}}
\def\eps{\epsilon}
\def\pd#1#2{{\partial#1 \over \partial#2}}
\def\ad{\a\_{D}}
\def\apj{Astrophys. J.}
\def\bk{\par\noindent}
\def\vu{\bu}
\def\vr{\br}
\def\vR{\bR}
\def\vf{\bff}
\def\vg{\bg}
\def\vF{\bF}
\def\vn{\bn}
\def\mi{m\_{i}}
\def\qi{q\_{i}}
\def\gfs{(\gf)\&{2}}
\def\hnmo{\hat\n^{-1}}
\def\E{E(\vr_1,...,\vr\_{N})}
\def\El#1{E(#1\vr_1,...,#1\vr\_{N})}
\def\p{\partial}
\def\Fmn{F_{\m\n}}
\def\FMN{F^{\m\n}}
\def\Jm{J^{\m}}
\def\Am{A_{\m}}

\title{NON-LINEAR CONFORMALLY INVARIANT
GENERALIZATION OF THE POISSON EQUATION
 TO $D>2$ DIMENSIONS}
\author{Mordehai Milgrom}
\address{Department of condensed-matter physics,
Weizmann Institute, Rehovot 76100 Israel}
\maketitle
\begin{abstract}
I propound a non-linear generalization of the scalar-field Poisson
equation of the form
$ [(\f\upud{i}\f\ud{i})^{D/2-1}\f\upcd{k}]\cd{k}\propto\r$,
 in curved $D$
dimensional space.
 It is derivable from the Lagrangian density $L\&{D}=\lfd
-A\r\f$, with
$\lfd\propto-(\f\upud{i}\f\ud{i})^{D/2}$,
 and $\r$ the distribution of
sources. Specializing to Euclidean spaces, where the field equation is
$\div(\abs{\gf}\&{D-2}\gf)\propto \r$, I find that
 $\lfd$ is the only conformally invariant (CI) Lagrangian in $D$
dimensions, containing only first derivatives of $\f$, beside the free
Lagrangian $(\gf)^2$, which underlies the Laplace equation. When $\f$ is
coupled to the sources in the above manner, $L\&{D}$ is left as the only
CI Lagrangian. The symmetry is one's only recourse in solving this
non-linear theory for some nontrivial
configurations. Systems comprising $N$ point charges are special
and afford further application of the symmetry. In spite of the CI, the
energy function for such a system is not invariant under conformal
transformations of the charges positions. The anomalous transformation
properties of the
energy stem from effects of the self energies of the charges. It follows
from these
that the forces $\vF_i$ on the charges $\qi$ at positions $\vr_i$  must
satisfy certain constraints beside the vanishing of the net force and net
moment: For example $\sum\_{i}\vr_i\cdot\vF_i$ must equal
 some given function of
the charges. The constraints total $(D+1)(D+2)/2$, which tallies with
the dimension of the
conformal group in $D$ dimensions. Among other things I use all these to
derive exact expressions for the following quantities: 1. The general
two-point-charge force. 2. The full potential field for two opposite
charges $\pm q$. 3. The energy function and the forces in any three-body
configuration with zero total charge. 4. The few-body force for some
special configurations. 5. The virial theorem for an arbitrary, bound,
many-particle system relating the time-average kinetic energy to the
 particle charges.
I also discuss briefly multi-scalar theories, theories with higher
derivatives, and vector- and higher-form-potential theories.
\end{abstract}

\section{Introduction}
\setcounter{equation}{0}
\par
It is a well-known and well-used fact that
the Poisson equation, $\Delta\f\prop \r$, for the potential $\f$
 produced by sources $\r$,
 describes a conformally invariant (CI) theory in two dimensions:
 It is invariant under the angle-preserving coordinate transformations.
 In all dimensions it is linear in the field $\f$,
and thus describes a ``free'' (non-self-interacting) field.
 Many of the special features of the
 $D=2$ theory stem from its linearity, but many are underpinned by the
conformal invariance. The Poisson equation in $D>2$ dimensions is not CI
(while the Laplace equation is--see section IIB).
\par
The Poisson equation describes many physical problems in linear media
such as
electrostatics, magnetostatics, steady-state diffusion and other
potential flows in the presence of sources and sinks, and, of course,
Newtonian gravity. It can be generalized to
\beq \div[\m(\abs{\gf})\gf]\prop \r,  \label{futura} \eeq
to describe, for example, non-linear media with a response coefficient
(dielectric constant, permeability, diffusion coefficient, etc.)
that is a function of the field strength. An equation of this type, with
different forms of $\m(x)$, has been studied in different contexts.
 For example, as an effective-action approximation to Abelianized
QCD\cite{ap}, as a modification of Newtonian
gravity to replace the dark-matter hypothesis for galactic systems
\cite{bm}\cite{sol},
and in the context of non-linear composite media\cite{bluber}.
Some of the general properties of such theories are summarized and
extended in \cite{acc}. Equation(\ref{futura}) might serve as a model
for many other nonlinear phenomena, such as electrodynamics in very
strong fields.
\par
 Here I point out that with the special choice
of $\m(x)\prop x^{D-2}$ the theory is a natural generalization of
 the (linear) two-dimensional Poisson theory.
 The resulting non-linear theory is unique in certain regards.
 Foremost is its conformal invariance. This enables one to say much about
 the theory and its
solutions--much beyond what is possible for the general case.
The theory seems to be the only one derivable from a CI
action that contains only first derivatives of the potential,
with the source $\r$ coupled directly to the
 potential, i.e. with an interaction Lagrangian of the form $\r f(\f)$.

\par
In the modified dynamics discussed as an alternative
to dark matter, phenomenology requires
just this CI behavior in three dimensions in the limit of
very small $\abs{\gf}$ (see \cite{bm}\cite{sol}).  Our results here
apply then in the large-distance limit of this theory.
\par
In material media, non-linearities of the response coefficient appear
 at high values of $\abs{\gf}$. Our results might then apply
in the short-distance limit. So, for example, our results for point
 charges will be valid
when charges are very near each other, and those for the fields
 at short distances from the sources.
\par
The present theory  constitutes an instance of a highly non-linear
theory that can be solved for non-trivial configurations due to the
symmetry.
\par
I shall present two types of results: One concerns solutions for the
potential field for various charge distributions obtained by conformal
transformations from highly symmetric ones; this I do in section III,
  after discussing some general properties of the theory in section II.
\par
The other type of results concern systems of point charges.
The dynamics of these is governed by an energy function
that depends on the charges and their positions. It turns out that while
the theory is invariant under conformal transformations,
the energy--surprisingly perhaps--is not invariant under a conformal
transformation of the positions of the point charges.
This can be seen already in the two-dimensional case, which is exactly
solvable, where the energy of a system of charges $\qi$ at positions
$\vr_i$, is $E=\sum_{i\not =j}q_i q_jln\abs{\vr_i-\vr_j}$.
Under a dilatation $\vr_i\rar\l\vr_i$ we have
 $E\rar E+ln\l\sum_{i\not = j}q_i q_j=
 E+(1/2)ln\l[(\sum_{i}\qi)^a-\sum_{i}\qi^a]$, with $a=2$.
 In the $D$-dimensional, non-linear case we do not, in general,
 have a closed expression for the energy. Still, we shall see that
 the energy transforms under dilatations as $E\rar E+ Kln\l$,
with $K$ a function of the charges of the same form, with a value of
the power $a$ that depends on $D$. The non-trivial term in the
transformation law comes from the behavior of the self energies
of the charges under dilatation, including the fact that the self
energy of a charge is logarithmic in its size scale. All this is
rather transparent in the linear two-dimensional case. There is also
an appropriate transformation law of the energy under inversions--the
other conformal transformations (and of course, the energy is invariant
under translations and rotations, which do not affect the self energies
of charges). I discuss all this in
section IV. Some of the applications to calculating energies and forces
 are brought in section V. In section VI, I
discuss other field actions for $\f$, and demonstrate the uniqueness
of $L\&{D}$ as a CI Lagrangian. In section VII, I discuss multi-potential
 theories. In section VIII I generalize briefly to non-linear, CI
 extensions of Maxwellian electrodynamics in $D>4$ dimensions.
 In the last section I bring brief comments on possible
 connections with quantum field theory.

\section{General properties}
\setcounter{equation}{0}
\par
Via the equation
\beq \div\{[(\gf)^2]^{D/2-1}\gf\}=\ad G\r \label{feq} \eeq
 a charge distribution $\r(\vr)$ in $D$-dimensional, Euclidean
space gives rise to a potential $\f(\vr)$.
This field equation is derivable from the action
\beq \sd=\si+\sf\equiv -\int~\r\f\dDr
-{1\over D\ad G}\int [\gfs]^{D/2}\dDr.
\label{action} \eeq
Here, $\sf$ is the field action, $\si$ is the interaction action,
$G$ is a coupling constant, and $\ad=2(\pi)\&{D/2}/\Gamma(D/2)$
 is the $D$-dimensional solid angle, introduced here for convenience.
 For $G>0$, like charges attract, as in gravity; for $G<0$ they repel
each other, as in electrostatics.
The field equation has a unique solution in a volume $V$ bounded by
$\Sigma$ when either $\f$ or the normal component of
 $[(\gf)^2]^{D/2-1}\gf$ are dictated on $\Sigma$ (see e.g.
 \cite{sol}\cite{acc}).

\par
Two integral relations satisfied by solutions of the
field equation were derived in \cite{acc}.
 The first applies for our class of
theories when the total charge vanishes; it then tells us that
\beq {1\over \ad G}\int [\gfs]^{D/2}\dDr=-\int~\r\f\dDr.
\label{muyaret} \eeq
 The second relation is an explicit expression for the virial integral
 $\VV$:
\beq \VV\equiv\int~\r\vr\cdot\gf~\dDr
=(dG)\&{-1}\abs{GQ}^d\equiv\vd(Q)  \label{expow} \eeq
[$d\equiv D/(D-1)$], which follows by writing $\VV$ as a surface
integral at infinity.
 The virial--which is shown below to control the response of the
configuration's energy to rescaling--can
 then be written in terms of only the total charge
 of the system.
\par
While I shall work in Euclidean space with its specific conformal
 transformations, it is useful to formulate the problem for
  curved space. The covariant form
of the action is
\beq \sd=-\int~g^{1/2}\r\f\dDr
-{1\over D\ad G}\int g^{1/2}(\Gij\f\ud{i}\f\ud{j})^{D/2}\dDr,
\label{covaction} \eeq
giving rise to the field equation
\beq [(\Gij\f\ud{i}\f\ud{j})^{D/2-1}\Gkm\f\ud{k}]\cd{m}=\ad G\r.
\label{cogatiza} \eeq
Above, $\gij$ is the metric,  $\Gij$ its inverse, $g=\abs{det(\gij)}$,
 and summation over repeated indices is understood everywhere.
The density, $\r$, is defined so as to be a coordinate scalar: the charge
within a volume $V$ is $\inv g^{1/2}\r\dDr$. So, for example, for a
 system of point charges $\qi$ at $\vr_i$
\beq \r(\vr)=g^{-1/2}\sum_{i}\qi\d\&{D}(\vr-\vr_i). \label{mitata}\eeq
 Using usual derivatives instead of covariant ones eq.(\ref{cogatiza})
reads
\beq g^{-1/2}
 [g^{1/2}(\Gij\f\ud{i}\f\ud{j})^{D/2-1}\Gkm\f\ud{k}]\ud{m}=\ad G\r.
\label{cogapipa} \eeq
\par
The field stress tensor is defined as the functional derivative of the
 field action with respect to the metric; i.e.,
under a small change $\d\gij$ in the metric
\beq \d\sf=\oot\int g^{1/2}\d\gij\PPs^{ij}\dDr, \label{lamurta} \eeq
giving
 \beq \PPs^{ij}=-{1 \over D\ad G}(\f\upud{k}\f\ud{k})^{D/2}
(\Gij-D{\f\upud{i}\f\upud{j}\over \f\upud{m}\f\ud{m}}),
\label{poxitar}\eeq
which has a vanishing trace. (The metric does not appear in the
interaction part--because $g^{1/2}\r$ depends only on the charges
degrees of freedom--which, thus, does not contribute to $\PP$.)
 The tracelessness results from the
conformal invariance of the action $\sf$, as is well known (see below).
 For the Euclidean case the stress tensor becomes
 \beq \PP=-(D\ad G)^{-1}\abgf^D(1-D\vn\otimes\vn),  \label{potipal} \eeq
with $\vn=\gf/\abs{\gf}$.
\par
 In this flat case,
the stress tensor gives the force on any volume $V$, bounded by the
 surface $\Sigma$ on which $\r=0$ as
\beq \vF\equiv~-\inv\r\gf\dDr=~-\ins~\PP\cdot \ds. \label{jutida} \eeq
(Compare with the expression of the force as a surface integral in
\cite{bm}.)


\subsection{Conformal coordinate transformations}
\par
The crux of this paper is that the above theory is invariant under
conformal coordinate transformations in $D$-dimensional space.
These are the angle-preserving transformations $\vr\rar\vR$,
for which the metric transforms as
\beq \gij\rar\pd{r^k}{R^i}g_{km}\pd{r^m}{R^j}=\l^2(\vr)\gij,
 \label{zeugma} \eeq
corresponding to local rescaling of distances.
The Jacobian determinant of the transformation is thus
 $J=\abs{\p R/\p r}=\l^{-D}(\vr)$.
\par
In a flat (Euclidean) $D>2$ dimensional space
 the group of conformal coordinate transformations comprises
 the rigid transformations (translations, rotations, and reflections)
under which the metric does not change; dilatations (rescaling)
$\vr\rar\l^{-1}\vr$, with a constant $\l$, for which
 $\gij=\d_{ij}\rar\l^2\d_{ij}$; and inversions. Under an inversion
about a sphere of radius $a$ centered at
an arbitrary point $\vro$ a point $\vr$ is transformed to a point $\vR$
on the same ray issuing from $\vro$, with the geometric mean
of the distances of $\vr$ and $\vR$ from $\vro$ being $a$. Explicitly:
\beq \vr\rar\vR=\vro+{a^2\over\abs{\vr-\vro}^2}(\vr-\vro).
\label{pupul}\eeq
The Euclidean metric $\d_{ij}$ then transforms as
\beq \dij\rar J^{-1}_{ik}\d_{km}J^{-1}_{jm}=
{a^4\over\abs{\vR-\vro}^4}\dij=
 {\abs{\vr-\vro}^4\over a^4}\dij, \label{kukumaka} \eeq
where
\beq J_{ij}=
 \pd{R_i}{r_j}={a^2\over\abs{\vr-\vro}^2}(\d_{ij}-2n_i n_j).
\label{kutata} \eeq
Here, $\vn$ is a unit vector in the direction of $\vr-\vro$.
The matrix in brackets has eigenvalues 1 ($D-1$ degenerate) and -1
(non-degenerate).
The determinant of $J_{ij}$, in absolute value, is thus
\beq J={a^{2D}\over\abs{\vr-\vro}^{2D}}=
{\abs{\vR-\vro}^{2D}\over a^{2D}}.  \label{kuytar} \eeq
\par
All conformal transformations take spheres into spheres (hyperplanes
included as spheres of infinite radius).
\par
Since the scalar potential transforms as
 $\f(\vr)\rar \hat\f(\vR)=\f[\vr(\vR)]$, we have
 $\grad_r\f\rar\grad_R\f[\vr(\vR)]=(\pdline{\vr}{\vR})\grad_r\f$,
 from which follows that
\beq (\grad_R\hat\f)^2={a^4\over\abs{\vR-\vro}^4}(\grad_r\f)^2.
 \label{lokoloka} \eeq
\par
It is customary to use,  instead of pure inversions,
transformations of the form $P_{\bA}=I_0 T(\bA)I_0$,
 where $T(\bA)$ is a
 translation by a vector $\bA$, and $I_0$ is the inversion at the
origin about a sphere of unit length. These have certain advantages:
they are connected continuously to unity, and they
 bring the properties of the conformal group
 into better relief. I prefer, however,
 to use pure inversions in what follows,
as they are easier to handle: they are self inverse, and have a simpler
transformation Jacobian.


\subsection{Conformal invariance of the theory}
\par
If $\vr\rar\vR$ is a conformal coordinate transformation,
our action and the field equation are invariant under replacement of
$\f(\vr)$ by $\f[\vr(\vR)]$, of $\r(\vr)$ by $J^{-1}\r[\vr(\vR)]$,
and of the metric $\gij(\vr)$ by $\gij[\vr(\vR)]$ (and $\dDr$ by
$d\&{D}R$ in the action).
This can be checked by direct substitution, but is easier to see as
follows:
In a general metric space,
 a theory is conformally invariant if its
action is invariant under replacement, everywhere in the action, of
$\gij$ by $\z^2(\vr)\gij$ (and thus of $\Gij$ by $\z^{-2}\Gij$,
and of $g$ by $\z\&{2D}g$), and of $\r$ by $\z\&{-D}\r$ [because of the
factor $g^{-1/2}$ in the definition of $\r$--see eq.(\ref{mitata})]
It is evident from expression(\ref{covaction}) for the action, or from
the field equation(\ref{cogapipa}), that ours is indeed a conformal
 theory by this definition.
 This implies conformal invariance in the above sense, evident by
applying first a conformal coordinate transformation, under which
$\f(\vr)\rar\f[\vr(\vR)]$, $\r(\vr)\rar\r[\vr(\vR)]$,
and $\gij(\vr)\rar J^{-2/D}\gij[\vr(\vR)]$ [see eq.(\ref{zeugma})].
The action, being a coordinate scalar, is invariant.
 Now transform the metric
back by multiplying it by the conformal factor $\z^2=J^{2/D}$, and $\r$
by $\z\&{-D}$. The action
remains invariant by virtue of its CI. The net result is that
 the action is invariant under the transformation described at the head
of this sub-section.
\par
It follows from this that
 if $\f(\vr)$ solves the field equation for the source $\r(\vr)$ and
 metric $\gij(\vr)$, then $\hat\f(\vR)=\f[\vr(\vR)]$
 solves it for the source $\hat\r(\vR)=J^{-1}(\vr)\r[\vr(\vR)]$, with
the same metric $\gij[\vr(\vR)]$.
Clearly, equipotential surfaces are transformed into equipotential
surfaces. Also, field lines go to field lines,
because they are perpendicular to equipotential surfaces
and angles are preserved in the transformation.
Charges are preserved in the transformation; i.e., the total charge
in a certain volume is the same as the transformed charge
in the image of that volume.
\par
  The tracelesness of the
stress tensor follows by employing eq.(\ref{lamurta}) with
 $\d\sf=0$ for $\d\gij=\eps(\vr)\gij$, with $\eps$ an arbitrary,
 (infinitesimal) function.
\par
 The application of such conformal invariance is standard
in the linear, two-dimensional case (in electrostatics, in potential-flow
problems, etc.). In $D$ dimensions such application have
special value  because the symmetry is our only recourse in
solving some of the problems in this strongly non-linear theory, as I
do in sections III-V.
\par
 The covariant Laplace (free) action,
$propto\int g^{1/2}\Gij\f\ud{i}\f\ud{j}\dDr$, is not CI in the above
sense,
but can be made so by adding to the above
 action a term proportional to $R\f$, with $R$ the scalar
 curvature, and taking
 $\f$ to have non-zero dimension, so that it transforms as
$\f\rar\l^{-(D/2-1)}\f$ (see e.g. \cite{bd}). The Euclidean Laplace
theory thus becomes CI with $\f$ of dimension $D/2-1$ (as the term with
the curvature vanishes), but then
the CI of the interaction term $\int g^{1/2}\r\f$ is lost. What is
 special about our theory, and what leads to the applications below,
is the fact that it is a CI theory in the presence of sources.
\par
Hereafter I confine myself to the Euclidean case. In curved spaces that
are conformally flat, such as maximally symmetric spaces, conformal
invariance implies the existence of coordinates in which the theory
takes the Euclidean form, with $\gij$ replaced by $\d_{ij}$ everywhere.

\subsection{Asymptotic behavior of the potential}
\par
If the sources $\r$ are contained within a finite volume, and sum up to a
total charge $Q\not = 0$, the field  becomes
radial at infinity, and, applying Gauss's theorem to the field equation
for a sphere of a large radius, $r$, we find asymptotically
\beq \gf\approx s(QG)\abs{GQ}^{1/(D-1)}
r\&{-1}\vn\_{r}. \label{asym} \eeq
 Here $s(x)=sign(x)$, and $\vn\_{r}$
is an out-pointing, radial unit vector. The potential is then logarithmic
for any dimension.
 When $Q=0$, the
 asymptotic behavior of $\f$ is determined by higher multipoles.
Typically, a dipole potential dominates asymptotically, and has the
form $\f\prop z/r^2$ (see below) with $z$ the axis along the dipole.
 Outside a spherical distribution of zero total charge
the field vanishes.


\subsection{Scaling properties}
\par
The field equation enjoys a two-parameter family of scaling
invariances: If $\f(\vr)$ solves the equation for a source $\r(\vr)$,
then, for any two constants $a$ and $b$,
$\hat\f(\vr)\equiv a^{-1}\abs{a/b}^{d}\f(b\vr)$ solves it for
$\hat\r(\vr)=a\r(b\vr)$, where $d\equiv D/(D-1)$.
When $b^D=a>0$, so that the total charge remains the same, the scaled
potential is $\hat\f(\vr)=\f(b\vr)$.
\par
It follows then that the potential, the electric field, the forces,
etc. scale simply with charge: If $\r\rar a\r$, than
 $\f\rar s(a)\abs{a}^{1/(D-1)}\f$, and forces (which scale like $q\gf$)
$\vF\rar\abs{a}^d\vF$. These quantities
 also scale with system size.

\section{ EXACT SOLUTIONS FOR THE FIELD}
\setcounter{equation}{0}
\par
Only few charge configurations with exact solutions are known
for the general case with an arbitrary form of $\m(x)$
in eq.(\ref{futura}) \cite{acc}.
In particular, there is a closed-form solution for any configuration
with one of the $D$
one-dimensional symmetries: plane-parallel, cylindrical,...,
 spherical: By applying Gauss's theorem we get for the present theory
\beq \deriv{\f}{R}\propto R^{-s/(D-1)}, \label{nununu} \eeq
where $R$ is the only coordinate on which $\f$ depends, and $s=0$
for the plane-parallel case, $s=1$ for the cylindrical case, etc..
 For a spherical system $s=D-1$, and we have
\beq \deriv{\f}{r}={[Q(r)]^{1/(D-1)}\over r}, \label{fututata} \eeq
where $Q(r)$ is the accumulated charge at spherical radius $r$
(here and in the rest of the section I use $G=1$).
 In the plane-parallel case
\beq \deriv{\f}{z}=[\ad \Sigma(z)/2]^{1/(D-1)}, \label{sutareda} \eeq
where $\Sigma(z)$ is the total surface density to the left (small-$z$)
of $z$ minus that to its right.
\par
These solutions, and others, may be used
 to generate new ones by applying
conformal transformations to the corresponding charge configuration.
Some examples follow.

\subsection{Two opposite point charges $\pm q$ at $\vr_1$ and $\vr_2$}
\par
Start
with a point charge $q>0$ at $\vr_1$ and a spherical shell evenly charged
with charge $-q$, centered at $\vr_1$, and having a very large
radius (infinite in the limit).
Upon inversion about a sphere of radius $a=\abs{\vr_1-\vr_2}$
 centered at $\vr_2$ the large spherical shell is transformed
 into a point charge $-q$ at $\vr_2$, and the charge $q$
stays at $\vr_1$.
 The potential for the original, spherically
symmetric system, is $\f(\vr)= q^{1/(D-1)}ln\abs{\vr-\vr_1}$ inside the
spherical shell, and $\f=0$ outside.
 It transforms into
\beq \f(\vr)=q^{1/(D-1)}ln{\abs{\vr-\vr_1}\over \abs{\vr-\vr_2}},
\label{julamop} \eeq
(after subtraction of the constant
$q^{1/(D-1)}ln\abs{\vr_2-\vr_1}$); this applies everywhere.
Interestingly, this potential is just the sum of the potentials of the
 two individual charges.
 This happens to be the case only for two opposite
 charges. It holds neither for two charges that are not opposite, or for
more then two charges.

\subsection{The pure-dipole field}
\par
Asymptotically, at $r\gg \ell$, where $\ell$ is the dipole separation,
the potential in eq.(\ref{julamop}) becomes
\beq \f\approx -q^{1/(D-1)}\ell{z\over r^2},  \label{mupkuta} \eeq
where $z$ is the dipole axis (positive charge
to the positive-$z$ side).
This is potential for a pure dipole of strength
 $q\ell^{D-1}$. It describes the field
everywhere in the limit $\ell\rar 0$ with $q\ell^{(D-1)}$ constant.
(For $D>2$, a standard dipole with $\ell\rar 0$ and $q\ell$ finite does
not contribute to the dipole field, due to self-screening effects.)
The pure dipole potential is a vacuum solution of the field equation that
is obtained from another vacuum solution: a constant-gradient field;
the latter has the potential $\f\prop z$ which transforms into $z/r^2$.
\par
The dipole field has a field strength, $\abgf\prop r^{-2}$,
that depends only on $r$--not on the angular coordinates. This is a
well-noticed property of the dipole field in two dimensions.
Here it follows directly from the transformation law(\ref{lokoloka})
 for $\abs{\gf}$, and the fact that the dipole field is obtained
by inversion from a constant-gradient field.
\par
 For a bounded density distribution of a vanishing total charge
the asymptotic behavior of the field is, generically, dominated
by a dipole field $Az/r^2$. I have not been able to express
$A$ as a functional of the density distribution.


\subsection{Point charge in the presence of a grounded sphere}
\par
 If, in the above example, we take the charged spherical shell
to have a finite radius, we end up with
 a point charge $q$ in the presence of
 a grounded sphere (as the potential on the original sphere vanishes).
 By a proper choice of the inversion radius and center,
the image point charge falls inside, or outside, the image sphere.
In the first case the potential inside the sphere is that of
two opposite charges, and vanishes outside; the tables are turned when
the charge falls outside the sphere.
The charge distribution on the grounded sphere is then easily
determined.

\subsection{Two oppositely charged spheres}
\par
 More generally, starting from two oppositely charged ($\pm q$)
concentric spheres (for which the potential is constant in the innermost
 and in the outermost regions, and is $q^{1/(D-1)}ln~r$ in between)
we get the potential field of two oppositely charged,
equipotential spheres of any sizes, either nested or detached.
When the spheres are nested, the potentials in the inner, and in the
 outer part are still constants; in between it is of the
 form(\ref{julamop}). When the spheres are detached the potential
is constant inside the spheres, and is of the form(\ref{julamop})
 outside.
\par
 Starting with two parallel hyperplanes charged with a constant
surface density $\pm\Sigma$ (between which $\gf$ is constant),
 and inverting about a point halfway between
the planes, we obtain two oppositely charged, tangent spheres. The
 potential vanishes inside the spheres; outside we
have an exact dipole potential. The charge distribution
 on each sphere--straightforwardly calculated--diverges
at the origin and together the charges give a dipole of finite strength.


\subsection{Some general comments on potential fields}
\par
Since equipotential surfaces remain so when transformed, and since
spheres go to spheres, the equipotential surfaces in all the above
 examples are spheres (all tangent in the case of a point dipole).
The field lines are all circles, being images of circles or straight
lines. For example, for a finite-separation dipole
 the field lines are all the circles going through the two charges.
\par
There are constraints on the field that can be deduced
even when the full field cannot be calculated. As an example consider
a charge distribution that lies on a circle (with $Q=0$).
 The field $\gf$ at any point $\vr$ must be tangent
to any sphere, of any dimension, containing $\vr$ and the circle (because
the sphere can be transformed into a plane by inversion about a point on
 it). This provides some information on the field of any
 three-point-charge configuration, or on that of a square quadropole.
\par
Other vacuum solutions of the field equation can be formed by starting
from the known, exact solutions of one dimensional symmetry
(uniformly charged, one-dimensional
wire, two-dimensional plane, etc.). As an example take
a one-dimensional wire in three dimensions with a constant line
density $\sigma$. Working in cylindrical coordinates $R,z$ we write
the potential as $\f=(8 \sigma)^{1/2}R^{1/2}$. Inverting about
a point off the wire will give the field for certain
 ring-plus-point-charge configurations.
Inverting about a point on the wire gives a configuration whose
vacuum solution is $\f\prop R^{1/2}(R^2+z^2)^{-1/2}$. This corresponds
to a charge density $\hat\sigma(z)\prop z^{-2}$ (and there appears an
infinite opposite charge at the origin to compensate the infinite charge
of the wire).
\par
 For a general charge distribution,
the field near an arbitrary point, $\vr$, away from charges,
is conformally related to
an asymptotic field: Invert about a very small sphere devoid of charges,
and centered at $\vr$. All the charges are transformed
into the small sphere, and the $\vr$ goes to infinity. The asymptotic
field of the new configuration is related to the field near $\vr$
in the original configuration. If $\gf(\vr)\not = 0$, the image
asymptotic field is dominated by a dipole term. In an opposite example,
look at the field near the mid-point between two equal point charges,
 where $\gf=0$; the inverted configuration is a quadropole, with a point
charge $-2q$ flanked by two symmetric charges $q$; the asymptotic
field decreases faster than a dipole field. I have not been able to
determine this asymptotic behavior.


\section{MANY POINT CHARGES--GENERAL CONSTRAINTS}
\setcounter{equation}{0}
\par
$N$-point-charge configurations--comprising bodies
 whose extent is much smaller than their separations--afford
further application of the conformal symmetry.
Take then a system made of $N$ point charges $q_1,...,q\_{N}$ at
$\vr_1,...,\vr\_{N}$, respectively.
The information on the dynamics of the system is encapsuled
in the energy function $E(\vr_1,...,\vr\_{N},q\_1,...q\_n)$.
The energy of a charge distribution may be taken as $-S$.
This converges at infinity
 only when the total charge vanishes. We can still
use it when $Q\not = 0$, provided only its changes under change
of configuration are needed:
 only the energies of configurations with the
 same total charge can be compared (see \cite{bm}).
Changes in $E$ may be calculated as changes in $-S$. An infinitesimal
 change $\d\r$
in the charge distribution with no net change in the total charge
($\int\d\r\dDr=0$) thus produces a change $\int\f\d\r\dDr$ in the
 energy ($\d\r$ also induces an increment of $\f$ but the field equation
implies that this does not contribute to the increment of the action).
 The $N$-point-charge energy,
 $\E$, (suppressing the $q$ variables) is a special case giving the
 relative energies of the $N$ point charges in different configurations:
we are only interested in comparing $E$ values as the charges are
moved rigidly to different positions.
The energy also diverges (logarithmically) near point charges, but these
divergences can be subtracted as self energies. I, in fact, treat point
charges as small but finite bodies, and need not be concerned with such
divergences.
\par
The force on the $i$'th charge is given by
\beq  \vF_i=-\pd{E}{\vr_i}.   \label {enrgata} \eeq
\par
 I now derive an expression for a virial integral defined in
 eq.(\ref{expow}) that involves only the
positions of the charges and the net forces on them, but is oblivious
to internal forces and structure (the above virial involves
integration inside the charges). To this end consider the contribution,
 $\VV_i$, of the $i$th body occupying the small volume $\vi$:
 $\VV_i=\int_{\vi}\r\vr\cdot\gf\dDr$. Write for $\vr$ within the body,
$\vr=\vr_i+\vh$, where $\vr_i$ is the center of charge,
 $\vh$ is small and will be taken to 0
in the limit. We can then write
\beq \VV_i=-\vr_i\cdot\vF_i+\int_{\vi}
\r\vh\cdot\gf\dDh,  \label{bnbn}\eeq
where $\vF_i$ is the net force on the body.
The second term in eq.(\ref{bnbn}) does not necessarily vanish in the
 limit $\vh\rar 0$, because $\gf$ inside the body
diverges in this limit.
Let $\gf_i(\vr)$ be the field produced by the body when it is the only
one present. Write
 $\gf(\vr)=\gf_i(\vr)+\grad \kappa(\vr)$, where $\kappa$ is the
increment in the potential due to the presence of the other bodies
in the system. (Because of the non-linearity $\kappa$ is not just
 the field produced by all the other charges.)
 In the limit $\vh\rar 0$, $\gf_i$ diverges like
$\abs{\vh}\&{-1}$ (from the scaling properties)
  but $\grad\kappa$ remains finite. Thus,
$\int_{\vi}\r\vh\cdot\grad\kappa\dDh$ vanishes in the limit,
 and we are left
with $\int_{\vi}\r\vh\cdot\gf_i\dDh$. This
 is just the virial defined for the $i$th body
when it is isolated. Thus, from eq.(\ref{expow}),
\beq \int_{\vi}\r\vh\cdot\gf\dDh
\rar \vd(\qi)\equiv(dG)\&{-1}\abs{G\qi}^d. \label{expowi} \eeq
 Putting all the
above together we finally get an expression for the reduced virial
\beq\VV_r\equiv -\sum\_{i}\vr_i\cdot\vF_i=\vd(Q)-\sum_i\vd(\qi)
=(dG)\&{-1}\abs{G}^d(\abs{Q}^d-\sum\_{i}\abs{\qi}^d).
  \label{expowii} \eeq
\par
Note that the limit of a point charge is gotten from a finite charge
distribution $\r(\vh)$ by taking the limit $\l\rar 0$ of
 $\l\&{-D}\r(\vh/\l)$; this has enabled us to take the limit of
$\int_{\vi}\r\vh\cdot\grad\kappa\dDh$ to 0. Point-like
 bodies of finite, higher multipoles (dipole, quadropole, etc.) cannot
be described in this way (for example, for a point dipole
$\int_{\vi}\r\vh\dDh$ remains finite in the limit).
Our ensuing results are not valid for such bodies.

\subsection{Scaling behavior of  the energy function}
\par
Expression(\ref{expowii}) implies then that
$ E_{\l}\equiv E(\l\vr_1,...,\l\vr_n)$ satisfies
\beq \pd{E_{\l}}{\l}=\sum_i\l^{-1}[\l\vr_i\cdot\vF_i(\l\vr)]=
\l^{-1}[\vd(Q)-\sum_i\vd(\qi)]. \label{ratowii} \eeq
Integrating over $\l$ between 1 and $\l$ we get an important
homogeneity property of $\E$:
\beq \El{\l}=\E+[\vd(Q)-\sum_i\vd(\qi)]ln~\l. \label{axulani} \eeq
\par
I shall now derive this transformation law of $E$ in a different way,
which illuminates better its origin, and which will be of further use
 below. This is based on the invariance of the theory under
space dilatations--a fact that also underlies the derivation of
eq.(\ref{expowii}).
\par
Consider first
 the change in the energy of an arbitrary charge distribution
$\r(\vr)$ under a dilatation:
 $\r(\vr)\rar \r_{\l}(\vr)=\l^{-D}\r(\vr/\l)$.
 In light of the scaling laws described in (IIC),
$\f(\vr)\rar\f_{\l}(\vr)=\f(\vr/\l)$.
Integrating $\d E/\d\l=\int~\f_{\l}\pd{\r_{\l}}{\l}~\dDr$ between 1 and
$\l$ gives
\beq E(\r_{\l})-E(\r)=ln~\l\int~\r(\vr)\vr\cdot\gf~\dDr=\vd(Q)ln~\l,
\label{zumalala} \eeq
where $Q$ is the total charge, and I have used eq.(\ref{expow})
for the virial integral.
This is the result that underlies all our findings below.
It says that the virial integral is the single system parameter that
determines the variations in energy under scaling transformations.
\par
What is the change in $E$ when the charges are
moved from positions $\vr_i$ to positions $\l\vr_i$?
This can be achieved in two steps: First
apply a space dilatation to the charge distribution.
The centers of charge are then moved to the
new positions. But also, the charges themselves (taken as very small
but finite bodies) are dilated by the same factor; this is more than
we want, as we need to move the charges {\it rigidly} to the new
 positions. After this first step we have from eq.(\ref{zumalala})
\beq \tilde\El{\l}=\E+\vd(Q)ln~\l,
 \label{polala} \eeq
where $\tilde E$ is the energy of the dilated charges at their new
configuration.
\par
In the second step we dilate each charge separately by the inverse
factor to bring the configuration to the desired one.
The energy change in the second step can be calculated  in
the limit of very small size for the charges. It
 is then the sum of changes due to separate dilation
of the individual charges. This cannot be done when the bodies are not
much smaller than their separations, and, as before, does not apply if
the bodies have finite, higher multipoles.
 Using eq.(\ref{zumalala}) again for the individual charges
 yields the a change $\Delta E=-\ln\l\sum_i~\vd(\qi)$. Putting the two
 together we get eq.(\ref{axulani}).
The non-trivial transformation properties of the energy
function under $\vr_i\rar\l\vr_i$ (even when $Q=0$) thus
 have to do with the transformation of the self-energies of the point
 charges.
\par
Invariance under translations implies that $E$ must be a function
of only differences of $\vr_i$, such that $E(\vr_i+\ba)=E(\vr_i)$.
The derivative of this with respect to $\ba$ gives
\beq \vF\equiv\sum_i\vF_i=0. \label{mutacut} \eeq
Similarly, invariance under rotations
 implies that $E$ depends only on scalars, and that the total moment
on the system must vanish
\beq \M\equiv\sum_i\vr_i\otimes\vF_i-\vF_i\otimes\vr_i=0.
 \label{gutmara}  \eeq


\subsection{Behavior of the energy under inversions}
\par
What does inversion invariance tell us about how $E$
changes under inversions, namely, when moving $\qi$ rigidly from
 $\vr_i$ to $\vR_i$ according to eq.(\ref{pupul})?
We saw above that if the inversion is not to produce a new charge at the
center we must start with
a total charge $Q=0$; when $Q\not =0$ we may annul it by
 putting a charge $-Q$ at infinity.
Then, from conformal invariance, the energy (action) is
 conserved under a inversion transformation of the charge distribution.
As in the case of dilatations, such a transformation
 does not just move the charges to
 their new positions; it also transforms their inner structure; how?
When the charges are of very small size their shape change
is determined by the first derivatives $\pdline{\vR}{\vr}$,
[eq.(\ref{kutata}) with $\vr$ and $\vR$ interchanged].
This describes a reflection about a hyperplane perpendicular to
$\vn$ through the body's center,
 and a dilatation by a factor
${a^2/\abs{\vr_i-\vro}^2}=R^2_{i0}/a^2$, with $\vR_{i0}=\vR_i-\vro$.
 As before, bringing the charges back to their
 original size changes the total energy, which leads to

\beq E(\vR_1,...,\vR\_{N})=\E-\sum_i\vd(\qi)ln\left(
{R^2_{i0}\over a^2}\right). \label{kupalula} \eeq
Again, the fact that $E$ is not
 invariant under inversion of the positions results from
the effect on the self-energy of the charges.
\par
The derivatives of eq.(\ref{kupalula}) with respect to $a^2$ and to
 $\vro$, at fixed $\vR_i$, give sum relations for the forces.
The first gives eq.(\ref{expowii}) for the
reduced virial, again. The second gives the
vector sum relation
\beq -a^2\sum_i\vF_i+
\sum_i r^2_{i0}\vF_i-2\sum_i(\vr_{i0}\cdot\vF_i)\vr_{i0}+2\sum_i\vd(\qi)
\vr_{i0}=0.  \label{ktuplapa} \eeq
Equation (\ref{ktuplapa}) holds for the forces $\vF_i$ to which
the point charges are subject when at $\vr_i$, for all values
of $a$ and  $\vro$. Separating the dependence on $\vr$ and $\vro$
this equation can be written as
\beq \bI-a^2\vF+\vro^2(1-2\vn\otimes\vn)\cdot\vF
+2V\vro+2\M\cdot\vro=0,  \label{kukatara} \eeq
where $V\equiv \sum_i[\vr_i\cdot\vF_i-\vd(\qi)]$, $\vn=\vro/\abs{\vro}$,
 and
\beq \bI\equiv
\sum_i r^2_i\vF_i-2\sum_i(\vr_i\cdot\vF_i)\vr_i+2\sum_i\vd(\qi)
\vr_i.  \label{klamispa} \eeq
\par
 For eq.(\ref{kukatara}) to hold for any $\vro$ and $a$
 we must have separately
$\vF=0$, $\M=0$, $V=0$,  and the new
sum  relation $\bI=0$.
 The number of such relations totals $D+D(D-1)/2+1+D=(D+1)(D+2)/2$--and
tallies with the dimension of the conformal group.
 The first three relations were
 derived above from transformation properties of $E$ under translations,
rotations, and dilatations, respectively.
 Now we see that they all follow solely from
the transformation properties under inversions. This need not be
surprising: locally, all the former transformations can be obtained
from combinations of inversions.
 Any set of points in a finite volume can be translated,
rotated, or dilated by using a succession of inversions alone.
\par
How do the forces on the charges transform under inversion?
If $\vF_i$ and $\vF^*_i$ are the forces in the old and new positions,
respectively, then, from eq.(\ref{kupalula}), and choosing the origin
at the inversion point ($\vro=0$)
\beq\vF^*_i=-\pd{E(\vR_1,...,\vR\_{N})}{\vR_i}=\pd{\vr_i}{\vR_i}\cdot
\vF_i+2\vd(\qi){\vR_i\over R^2_i}. \label{locurat} \eeq


\section{N-POINT-CHARGE CONFIGURATIONS--SOME APPLICATIONS}
\setcounter{equation}{0}
Energy functions and forces for certain $N$ point charge configurations
can be calculated with the use of the results in the previous section.

\subsection{The two-body system}.
\par
 The force, $F(\qo,\qt,\ell)$,
 between two point charges $\qo and~\qt$,
 a distance $\ell$ apart ($F$ is positive for attraction) can be obtained
from expression (\ref{expowii}) for the reduced virial:
Here $\vF_1=-\vF_2=\vF$, and the vanishing of the moment implies
that $\vF$ points from one charge to the other, so that
$\VV_r=F\ell$, and we can write
\beq F={s(G)\over\ell}[\vd(\qo+\qt)-\vd(\qo)-\vd(\qt)]={s(G)\over \ell}
d\&{-1}\abs{G}^{d-1}(\abs{\qo+\qt}^d-\abs{\qo}^d-\abs{\qt}^d)
  \label{forcof} \eeq
[$d\equiv D/(D-1)$].
This result was derived in \cite{lsy} for the three-dimensional case
 in a roundabout manner. (For $D=2$ expression (\ref{forcof})
 reduces to the standard
two-dimensional linear-medium result $F=G\qo\qt/\ell$.)
 For example, for two equal charges $\qo=\qt=q$,
 $F=2s(G)\ell^{-1}d^{-1}\abs{G}^{d-1}\abs{q}^d(2^{d-1}-1)$. For opposite
charges: $\qo=-\qt=q$,
 $F=-2s(G)\ell^{-1}d^{-1}\abs{G}^{d-1}\abs{q}^d$, also to be
 gotten from the force-transformation law (\ref{locurat})
 starting with one charge at infinity, and hence $\vF^*=0$.
\par
The energy function in the two-body case is
\beq E(\vr_1,\vr_2,q_1,q_2)=\b_{12}ln\abs{\vr_1-\vr_2},
\label{futukap}  \eeq
where
\beq \b_{ij}\equiv\vd(q_i+q_j)-\vd(q_i)-\vd(q_j). \label{lotiaga} \eeq

\subsection{The three-body system with vanishing total charge}.
\par
Consider three charges $\qi$ at $\vr_i$,
 with $q_1+q_2+q_3=0$.
Perform an inversion with one of the $\vr_i$s as center, say $\vr_3$.
Then, $q_3$ is transformed to infinity, and $q_{1,2}$ are transformed
to $\vR_{1,2}$. The force on $q_1$, say, can be calculated in the
new configuration from the two-body force formula eq.(\ref{forcof}).
 From this the force $\vF_1$ in the original three-body
configuration is calculated by employing the force-transformation law
 eq.(\ref{locurat}) to obtain
\beq \vF_1=\b_{12}{\vr_2-\vr_1 \over r^2_{12}}
+\b_{13}{\vr_3-\vr_1 \over r^2_{13}}.
\label{threefa} \eeq
(This could also be derived from the above constraints on the forces
$\vF=0,~\M=0,~V=0,~\bI=0$.)
Interestingly, the force is the sum of the two forces that
would have been exerted by $q_2$ and $q_3$ separately, the non-linearity
notwithstanding.
Integrating eq.(\ref{threefa}) over $\vr_1$ we get the explicit form
of the three-body energy function (for the zero-total-charge case):
\beq E(\vr_1,\vr_2,\vr_3)=ln[r^{\b_{12}}_{12}
 r^{\b_{13}}_{13}  r^{\b_{23}}_{23}]. \label{threecu} \eeq


\subsection{The virial theorem}.
\par
I now derive the analogue of the standard  virial theorem,
  $\av {E_k}=-\av{ E_p}/2$, relating the mean kinetic energy and mean
potential energy of an $N$-body system held together by Newtonian
gravity in three dimensions.
Consider a bound system made of any number of
 point charges $\qi$, of masses $\mi$, moving under the sole
influence of the $\f$ field they produce (other forces may act inside
each body to hold it together).
The center-of-mass acceleration of each body
$\ddot{\vr_i}=\vF_i/\mi$, with the $\vF_i$ satisfying eq.(\ref{expowii}).
Now
 \beq \VV_r=-\sum_i\vr_i\cdot\vF_i= -\sum\_{i} \mi\vr_i\cdot\ddot{\vr_i}=
 -{1\over 2}{d^2\over dt^2}
 [\sum\_{i} \mi \vr_i^2]+\sum\_{i} \mi\dot{\vr_i}^2.
 \label{hayr} \eeq
The first term vanishes in the stationary case (or its long-time average
vanishes for a general bound system); the second term
equals twice the kinetic energy
 $E_k=M\av{\bV^2}/2$, where M is the total mass of the
system, and
$\av{\bV^2}$ is the mean square velocity. Together
with eq.(\ref{expowii}) this finally gives the desired virial theorem
 \beq 2E_k=M\av{\bV^2}=\vd(Q)-\sum_i\vd(\qi)=
(dG)\&{-1}\abs{G}^d(\abs{Q}^d
-\sum\_{i}\abs{\qi}^d),  \label{exkouti} \eeq
by which the mean-square velocity depends only on the charges.
This is exact for a stationary system;
 for a general bound system the long-time average of the
 left-hand side has to be taken.
\par
This relation has been used to estimate the mass of large-scale,
cosmological, galaxy filaments, which are approximately two-dimensional
Newtonian systems\cite{elt}. It has also been used to
determine the masses of roundish galaxies, in the modified dynamics,
  from their observed velocity dispersions,
 when typical accelerations in
them are very law (see \cite{lsy}\cite{sd} and references therein).
 In these cases, where $\qi$ are the constituent masses,
and where $N\gg 1$, $\sum\abs{\qi}^d$ can be neglected as it is smaller
than $\abs{Q}^d=M^d$ by a factor $\sim N^{-1/(D-1)}$. In the limit
 $N\rar\infty$
\beq \av{\bV^2}={D-1\over D}(GM)^{1/(D-1)}. \label{gutiasta} \eeq


\subsection{Symmetric configurations}
\par
 The force on bodies in some symmetric configurations can be calculated
from expression (\ref{expowii}):
Consider a configuration comprising a charge $q_0$ at the center, and
$n$ equal charges, $q$, at positions $\vr_i$
 that are equivalent with respect
 to the center--equivalent in the sense that each of the
 points, $\vr_i$, can be interchanged with
  any other by an element of the symmetry group of the system, $H$: a
 rotation, a reflection about the center, a reflection about
some hyperplane through the center, or combinations thereof,
 that is also a symmetry of the system. Examples are
 the corners of a rectangular hyper-box, the vertices
 of any perfect solid of dimension
$D$ or less (such a s a perfect polygon, any hyper-cube, etc.),
the vertices of polygonal prisms of different types, etc..
Clearly, the $q$-charges
 are then all at the same distance from the center, call
it $r$. Also, they are subject to equivalent forces $\vF_i$; to wit, each
 of the $\vF_i$s can be transformed to any other by  an element of the
symmetry group. In particular,
the radial components of these forces
$(\vF\cdot\vr)_i$ are all equal, because scalars are invariant under
the point group. (The force on the charge at the center vanishes.)
This common value can be deduced from eq.(\ref{expowii}) (the symmetry
automatically insures that $\vF=0,~\M=0$, and $\bI=0$):
\beq \vF\cdot\vr=-{1\over n}[\vd(q_0+nq)-\vd(q_0)]+\vd(q).
  \label{yampala} \eeq
When the forces are radial--as when the point are the vertices of a
 perfect solid--the full force is obtained since in this case
$\vF\cdot\vr=-Fr$ ($F$ is positive when $\vF$ acts towards the
 center).
\par
In the limit $n\rar\infty,~~nq\rar Q$, eq.(\ref{yampala}) gives  the
 force on the elements of
 a spherical shell of any dimension smaller than
 $D$ (e.g. a ring)
 having radius $r$, total charge $Q$, and a charge $q_0$ at its
 center. The force-per-unit-charge on the shell is
\beq F={1\over Qr}[\vd(q_0+Q)-\vd(q_0)]. \label{kukatila} \eeq
\par
Additional results are described in appendix A.


\subsection{Point charges in the presence of conducting boundaries}
\par
When (equipotential) bodies of infinite conductivity are present, the full
conformal symmetry enjoyed by the $N$-charge problem is destroyed (the
boundaries remain in place when applying the transformation to the
 charges). So, for example, homogeneity is always lost, and the total
 force now does not vanish in general; rotational symmetry sbout an
arbitrary center is lost, an so the total moment does not vanish; etc..
 Some symmetry may however be left, in which case the corresponding
 identities are still valid. If the arrangement of conductors is invariant
 under translations in a certain direction, the component of the total
 force on the charges in this direction vanishes.
If the conductors are spherically symmetric about some center, the total
 moment on the charges with respect to this center vanishes, and so on.
An interesting example involves boundaries that are invariant to rescaling
about a certain point (taken at the origin). This happens when for every
point, $\vr$, on the boundary, $\l\vr$ is also on the boundary, for every
$\l\ge 0$--the boundary is an arbitrary-cross-section cone (including
a hyperplane, a corner, etc.). In this case, expression (\ref{expowii})
for the reduced virial holds with respect to the origin (but $\vI=0$ does
 not); $Q$ now is the
total charge including that on the conductors. When the conductors are
grounded they automatically take up a charge that makes $Q=0$.
Thus, for example, the force on a single charge $q$, in the presence of
an arbitrary, grounded, conic conductor is always subject to a force $\vF$
satisfying
 \beq \vF\cdot\vr=\vd(q).\label{olopala} \eeq
This can be extended to conductors in the shapes of discs or spherical
caps, as they can be transformed into half-planes by inversions.


\section{OTHER CONFORMAL ACTIONS?}
\setcounter{equation}{0}
\par
Rotation invariance dictates that an
action containing only first derivatives of $\f$ is a function of
$(\gf)^2$, and scale invariance further requires that it be of the form
 $S^p\propto\int[(\gf)^2]^p\dDr$. By themselves, all such actions are
 invariant under the scale transformation
 $\f(\vr)\rar \hat\f(\vr)=\l^{\a}\f(\l\vr)$,
 with $\a=(D/2p)-1$, namely, with $\f$ having conformal dimension $\a$.
 If it is to be fully
CI, $S^p$ must also be invariant under inversion at the origin:
$\f(\vr)\rar \hat\f(\vr)=(a/r)^{2\a}\f(a^2\vr/r^2)$.
 This can be shown not to be
the case unless either $p=1$ (the linear case), or $p=D/2$, which is
our action $\sf$. To see this
 note that starting with $\f\prop r^{-(D-2p)/(2p-1)}$, which is the
only spherical vacuum solution of the theory (beside $\f=const$),
 the corresponding $\hat\f$ is not a solution, unless $p=1,~D/2$.
 Similarly, if we start with $\f\prop z$, which is a
vacuum solution for all $S^p$, the corresponding $\hat\f$ is not, unless
$p=1,~D/2$.
\par
 Field theory lore has it that scale-invariant theories tend to be CI,
 but this is not a theorem (see e.g. \cite{pol}, and references therein).
The above nonlinear theories constitute counter-examples.
\par
 For the interaction action to be
invariant we need $\f$ to transform according to
$\f(\vr)\rar\f[\vR(\vr)]$, since $\r\dDr$ is invariant; i.e.,
$\f$ is then
of conformal dimension zero.
As explained in section IIB,
the Poisson action in $D>2$ is not
CI because for the free action to be invariant $\f$ has to have dimension
$D/2-1$. We are thus left with $\sd$
 as the only CI action containing only first derivatives in the field
part.
\par
Consider now actions with a field part containing higher derivatives
written for curved space
\beq S=-\int~g^{1/2} L(\Gij,\DD\f,...,\DD^k\f)\dDr
 -\int~g^{1/2}\r\f\dDr.  \label{kupion} \eeq
The field Lagrangian $L$ depends on covariant derivatives of $\f$:
$\f\cd{i_{1}}...\cd{i_{m}}$, $1\le m\le k$
(collectively designated $\DD^m\f$), and is a coordinate scalar.
 Scale invariance alone is
tantamount to the homogeneity requirement
\beq L(\Gij,\l\DD\f,...,\l^k\DD^k\f)=\l^D L(\Gij,\DD\f,...,\DD^k\f).
\label{jifatreq} \eeq
This is because coordinate invariance dictates that every
derivative in $L$ is contracted with another by
 one $\Gij$. As $\f$ has zero dimension,
 under dilatations, $\vr\rar\l^{-1}\vr$, an  $m$th covariant
derivative is multiplied
 by $\l^m$, which is the same as multiplying $\Gij$ by $\l^2$, and
conformal invariance tells us that $L(\l^2\Gij,...)=\l^DL(\Gij,...)$.
\par
Consider, hereafter,
 the Euclidean version of the action
(where $\gij$ is put to $\d_{ij}$, and $\DD$ to $\p$).
As stated above, scale invariance does not insure full CI.
 The quadratic Lagrangians--leading to linear field
equations--with $\f$ of zero scaling dimension:
$ L=\f\Delta^{D/2}\f$ (for even $D$) are CI.
 The non-linear actions with $\f$ of zero scaling dimensions are
probably not. In fact, I have not been able to find any that is
 ( without having a general proof that none is). For example, I have
 shown that all the Lagrangians, in even $D$ dimensions, of the form
\beq L=[(\gf)^2]^m(\Delta\f)^k,  \label{kutaref} \eeq
 with  $ m=D/2-k$, and with $k=1$ for $D\ge 4$; $k=2$ for $D>4$, or
$k\ge 3$ for $D\ge 2k$, which are scale invariant, are not CI.
\par
The homogeneity condition(\ref{jifatreq})
 implies that the field
equation necessarily has a vacuum, spherically symmetric solution
of the form $\f=Aln~r$.
The Euler-Lagrange equation can be written in the form
$\p_iJ_i=\r$, where $J_i$ is a vector that is a function of the
first $2k-1$ derivatives of $\f$, with the homogeneity property
[easily derived from eq.(\ref{jifatreq})]
\beq J_i(\l\p\f,...,\l^{2k-1}\p^{2k-1}\f)=\l^{D-1}J_i(\p\f,...,
\p^{2k-1}\f).  \label{cutilma} \eeq
When $\f$ depends only on the radial coordinate $r$, $\bJ$ has only
an $r$ component, and, from eq.(\ref{cutilma}), for $\f=Aln~r$,
$J_r=r^{-(D-1)}j(A)$. Since in the spherical case
$\p_iJ_i=r^{-(D-1)}\p_r[r^{D-1}J_r]$, clearly $\f=Aln~r$ is a solution.
The coefficient $A$ is determined via the Gauss theorem: $j(A)\prop Q$,
where $Q$ is the total charge at the center.
If $L$ is homogeneous in $\f$ of degree $\b$, then $J$ is homogeneous
 of degree $\b-1$ in $\f$, and then the coefficient $A$ is given by
 $A\prop sign(Q)\abs{Q}^{1/(\b-1)}$.
\par
The purely logarithmic potential outside a single spherical body is,
however, valid for only very specific density runs.
When the Lagrangian depends on derivatives up to
 the $k$th,  the generic, spherically
symmetric, vacuum solution is characterized by $2k$ constants, which
are determined by the exact density run in the central body.
 In general, the potential
diverges at large $r$ as a power in $r$. The notion of a point
charge is thus not useful as the external solution does not depend only
 on the total charge. We cannot even speak of ``the field of a point
 charge'' as this is not well defined.
 For example, the quadratic theory with $\f$ of zero dimensions, with
$L\prop\f\Delta^{D/2}\f+A\r\f$, with the field equation
$\Delta^{D/2}\f\prop\r$, has $D$ independent
 spherically symmetric vacuum solutions of the form
 $\f=const.,~ln~r,~r^{\pm\a}$, with $\a=2,4,,...,D-2$.


\section{MULTI-POTENTIAL THEORIES}
\setcounter{equation}{0}
\par
 The above theory is straightforwardly extended to describe $K$
(coupled) scalar potentials, $\f_a$, which couple to $K$ types of
charges with densities $\r_a$  ($a=1,...,K$).
The action is
\beq S= -\int\sum_a\r_a\f_a\dDr
-{1\over 2\ad }\int \LL(\a_1,...,\a_K)\dDr,
\label{bunion} \eeq
with $\a_a\equiv (\gf_a)^2$, and $G=1$.
Conformal invariance is now equivalent to homogeneity of $\LL$:
\beq \LL(\l\a_a)=\l^{D/2}\LL(\a_a).  \label{lumatifa} \eeq
(One could actually generalize further by taking
 $\a_{ab}=\gf_a\cdot\gf_b$ as variables, but I keep to the simpler form.)
The $K$ field equations are
\beq \div[\m_a(\a_1,...)\gf_a]=\ad \r_a, \label{feqjuta} \eeq
with $\m_a=\p\LL/\p\a_a$, $1\le a\le K$.
The spherical vacuum solution for a system with total charges $q_a$
is
\beq \f_a=s(q_a)Q_a^{1/(D-1)}ln~r,  \label{futerav} \eeq
with $Q_a\ge0$, giving $\a_a=(Q_a)^{2/(D-1)}r^{-2}$;
$s(q_a)Q_a$ may be viewed as the asymptotically observed charges.
They are determined from the actual charges, $q_a=\int\r_a$, as follows:
Inserting the above form of $\f_a$ in the Gauss theorem
$\m_a(\a_1,...)d\f_a/dr=q_a r^{-(D-1)}$, and making use of the
homogeneity property of the $\m_a$ (derived from that of $\LL$):
$\m(\l x_1,\l x_2,...)=\l^{(D/2-1)}\m_a(x_1, x_2,...)$, we get the
$K$ equations:
\beq \m_a[Q_1^{2/(D-1)},Q_2^{2/(D-1)},...]Q_a^{1/(D-1)}=\abs{q_a}.
\label{vutigat} \eeq
\par
All our results in the previous sections
 can be carried through {\it mutatis mutandis} to this,
more general, case.
The virial integral, which controls the change in system energy under
dilatations, can now be shown to be given by
\beq \VV\equiv\int\sum_a\r_a\vr\cdot\gf_a~\dDr={D-1\over 2}
\LL[Q_1^{2/(D-1)},Q_2^{2/(D-1)},...,Q_K^{2/(D-1)}]. \label{zuzulat} \eeq
Using the homogeneity of $\LL$, eq.(\ref{lumatifa}), by which
 $\LL(x_1,...,x_K)=(2/D)\sum_a x_a\m_a)$, and with
 eq.(\ref{vutigat}), we can also write
\beq \VV={D-1\over D}\sum_a\abs{q_a}Q_a^{1/(D-1)}. \label{sulamira} \eeq
\par
This expression for $\VV$ is to replace $\vd(q)$ in all our results
(remember that the $q_a$s and $Q_a$s are charges of different
types of the same body). For example, the powers $\b_{ij}$ appearing in
the two- and three-body energy functions defined in eq.(\ref{lotiaga})
 are now given by
\beq \b_{ij}={D-2\over 2}\{
\LL[\hat Q_1^{2/(D-1)},...]-
 \LL[(Q_1^i)^{2/(D-1)},...]-
\LL[(Q_1^j)^{2/(D-1)},...]\},   \label{lolipoa} \eeq
where $Q_a^i~~a=1,...,K$ are the ``asymptotic'' charges for the point
body $i$, and $\hat Q_a$ are those for the two bodies $i,j$ taken
together; i.e., as calculated from eq.(\ref{vutigat}) with the charges
$q_a=q_a^i+q_a^j$.
\par
As a special case, assume that the theory is invariant under rotations
in the internal space of $\f$s, i.e., under $\f_a\rar\hat\f_a=
O_{ab}\f_b$ where $O$ is an orthogonal matrix (the charges are then
rotated by the same matrix, leaving $\sum_a\r_a\f_a$ invariant).
This means that $\LL$ must be a function of $\sum_a\a_a$,
 and the required
homogeneity then dictates that
\beq \LL={2\over D}\left(\sum_a \a_a\right)^{D/2},  \label{miutari} \eeq
(the constant in front is chosen to match the single-field case).
\par
Equation(\ref{vutigat}) can now be solved to give the asymptotic charges
as
\beq Q_a=\abs{q_a}^{D-1}q^{-(D-2)},  \label{fikunia} \eeq
where $q\equiv \left(\sum_a q_a^2\right)^{1/2}$ is the root-mean-square
over all the charge types of a body.
The virial now takes the single-field form, only with the system's
single charge replaced by its root-mean-square over all the charge types:
$\VV=\vd(q)$.


\section{Vector and higher-form theories}
\setcounter{equation}{0}
\par
Maxwell's electromagnetism is governed by the action
\beq S=-{1\over 4}\int~g^{1/2} \FMN\Fmn\dDr
 +\int~g^{1/2}\Jm\Am\dDr.  \label{pokupion} \eeq
It describes the electromagnetic field
 $\Fmn\equiv A_{\n}{}_{,\m}-A_{\m}{}_{,\n}$,
 derived from the vector potential
$\Am$, in the presence of conserved currents $\Jm$.
 The theory is gauge invariant, and is conformally invariant in four
 dimensions only ($\FMN=F_{\a\b}g\&{\a\m}g\&{\b\n}$, and
 $\Jm=g^{-1/2}j^{\m}$, where the vector density $j^{\m}$ contains only
matter degrees of freedom, but not the metric).
 In a vein similar to that in our treatment of the scalar case
we can construct non-linear, CI generalizations for $D> 4$:
Take the field action to be
\beq S_f=-\int~g^{-1/2}\LL(\Fmn,g_{\m\n})\dDr,  \label{hutarat} \eeq
where $\LL$ is homogeneous of degree
$D/2$ in $F$: $\LL(\eta\Fmn,g_{\m\n})=\eta\&{D/2}\LL(\Fmn,g_{\m\n})$
 (and, of course,
is a coordinate scalar). It is CI because multiplying $g_{\m\n}$ by
$\l(\vr)$ in $\LL$ is tantamount to multiplying $\Fmn$ by $\l^{-1}(\vr)$
(every two lower-case indices in $\Fmn$ must be contracted using one
$g^{\m\n}$). A factor $\l\&{-D/2}$ is then pulled out of $\LL$ to
cancel the factor from $g^{1/2}$. In $D>4$-dimensions we may take
\beq  \LL\propto (\Fmn\FMN)^{D/4},  \label{rujataa} \eeq
which gives a CI, vector theory when coupled to the currents as above,
but there are others. For example, in eight dimensions
\beq  \LL\propto F_{\m\a}F^{\a\b}F_{\b\c}F^{\c\m}  \label{rudada} \eeq
is such a theory.
\par
More generally, we have linear, gauge-invariant, CI theories in even
$D$ dimensions involving an antisymmetric-tensor gauge potential
of rank $n=D/2-1$ (an $n$-form potential), $A_{\a\_{1},...,\a\_{n}}$
 The field tensor $H_{\a\_{1},...,\a\_{n+1}}$ is the totally
antisymmetrized derivative of $A$, and the current density is also an
$n$-form, in analogy with the Maxwellian case.
The field Lagrangian is $\LL\propto H_{\a\_{1},...,\a\_{n+1}}
H^{\a\_{1},...,\a\_{n+1}}$, and the interaction Lagrangian is
$A_{\a\_{1},...,\a\_{n}}J^{\a\_{1},...,\a\_{n}}$.
 In dimensions higher than $2(n+1)$  we
get a CI theory by taking $\LL(H)$ that is homogeneous of degree
$D/(n+1)$ in $H$, e.g.
 $\LL\propto (H_{\a\_{1},...,\a\_{n+1}}
H^{\a\_{1},...,\a\_{n+1}})^{D/2(n+1)}$.
There is no known linear, CI generalization in
 $D>2(n+1)$(see e.g. \cite{ds}).
\par
Specialize now to flat spaces, I find that,
as in the scalar case, the effect of rescaling on the
energy of some current distribution is given by
\beq E_{\l}-E=\VV~ln~\l, \label{zugatara} \eeq
where $E_{\l}$ is the energy of the rescaled current distribution.
Again, $\VV$ can be written as a surface integral. For example, for the
vector-potential case, where the rescaled current density is
$\l\&{-(D-1)}\Jm(\vr/\l)$, we have
\beq \VV=\int d\s\_{\n}~(r^{\n}\LL-2r^{\b}A_{\a}{}_{,\b}
\pd{\LL}{F_{\n}{}_{,\a}}). \label{gratiuta} \eeq
This can be shown to vanish for configurations with
spacially bounded currents (bounded in space and time in higher-$D$
 Minkowski spaces).
We do not have the appropriate analogues of point charges in the
scalar case, with finite $\VV$, for which
 to calculate and manipulate $N$-point energies.
The situation is akin to having, in the scalar case, a system of point
bodies of null charge but a finite higher multipole.
As explained in section IV, our treatment of point-charge systems
 does not carry to such systems.
\par
We can still employ the CI to create new solutions of the non-linear
theory from other known solutions. For example, the vector potential
 $\Am=(1/2)B_{\a\m}r^{\a}$, with $B_{\a\m}$ constant and
 antisymmetric, gives
a constant field $\Fmn=B_{\m\n}$, and is thus a vacuum solution of all
the above vector theories in $D$ dimensions. Inversion at the origin,
under which $\Am(\vr)\rar r^{-2}\Am(\vr/r^2)$ ($\Am$ is of
dimension one in these theories) gives
  $\hat\Am=(1/2)B_{\a\m}r^{\a}r^{-4}$, which must them also be a
 solution.

\section{DISCUSSION}
\setcounter{equation}{0}
\par
It was pointed out to me by David Kutasov (private communication) that
our results for the classical field theory
 evoke, and might well be related to, results known to hold in conformal
field theory. Perhaps there is such a conformal QFT
whose classical limit is our theory. Comparing with the definition
of a conformal QFT as given e.g. in \cite{gin}, to make such a connection
we will have to identify the so-termed ``quasi-primary'' fields of the
QFT with $e^{iq\f(\vr)}$, where $\f(\vr)$ is now a quantum field.
 The $N$-point-charge energy function is the classical limit of
 the correlation function of these quasi-primary fields
\beq \E=ln[\langle e^{iq_1\f(\vr_1)}\cdot...\cdot
e^{iq\_{N}\f(\vr\_{N})}\rangle],  \label{vutiata} \eeq
and the constraints I found such as (\ref{expowii})(\ref{ktuplapa})
are the Ward identities for the correlator. What I then proved amounts
 to showing that the (anomalous) dimension of the operator
 $e^{iq\f(\vr)}$ is $\vd(q)$.
\par
 In the vector- and higher-form-potential case we cannot
identify such an in finite set of "quasi-primary" fields.

\begin{acknowledgements}
I thank David Kutasov for most valuable suggestions and discussions,
and Jacob Bekenstein for helpful comments on the manuscript.
\end{acknowledgements}


\setcounter{section}{0}
\def\thesection{\Alph{section}}
\def\theequation{\Alph{section}.\arabic{equation}}

\section{SOME MORE EXAMPLES OF FORCE CALCULATIONS}
\setcounter{equation}{0}
\par
The forces on the charges can also be calculated for symmetric
configurations of the following type:
Start with the symmetric configuration described in section VD,
 with an even number of sites, $n$.
Put the charge at the center to 0. Place an equal number of positive and
negative charges $\pm q$ at the sites in such a way that all charges
of the same sign are equivalent:
 Every two charges of the same sign can be interchanged
by a point symmetry that does not mix charges of a different sign.
 The corners of a $3-D$ rectangular box, for example, can be decorated
in three inequivalent ways that satisfy the above. The vertices of
a perfect $4m$-polygon can be decorated in two ways: alternating
 charges, and in a two-pluses-two-minuses pattern.
Here again the forces on all charges, positive as negative,
 are equivalent
(can be transformed to each other by elements of the full, site point
group $H$).
 Again,
$\vF\cdot\vr$ takes the same value for all the charges. Using
 expression(\ref{expowii}) we get
\beq \vF\cdot\vr=\vd(q).\label{gupzala} \eeq
\par
A charge configuration of the above description results when we apply
the method of images to the problem of a charge $q$ in a region
bounded by two intersecting, grounded, $D-1$-dimensional hyperplanes.
 When the angle between the hyperplanes is $\a=\pi/m$,
the field in the region bounded by the planes is the same as that of
a system of images. This has $m$ pairs of charges $\pm q$ arranged
alternately on the vertices of a polygon in a configuration as above
(the polygon is not perfect but has edges of alternating lengths).
Equation (\ref{gupzala}) thus gives the radial force on the charge; it is
 a special case of eq.(\ref{olopala}).
\par
Now an example of the use of the constraint $\bI=0$, where $\bI$
is defined by eq.(\ref{klamispa}). Consider
a planar, perfect polygon of $n$ equal charges $q$ (a uniform
 ring in the limit $n\rar \infty$), and a charge $-nq$
on the symmetry axis of the polygon, a distance $\ell$ from the origin
at the polygon's center.
We want the force $F$ on the large charge (which acts along the axis),
 and the force $\vf$ on
the small charges.
 From $\bI=0$ we have
\beq F=-{2\vd(nq)\ell\over r^2+\ell^2}, \label{gulitara} \eeq
where $r$ is the radius of the polygon. The axial component of the force
on the small charges is $-F/n$, and the radial component, $f_r$
 is gotten from $V=0$:
\beq f_r=-r^{-1}\left[\vd(q)+{1\over n}
\vd(nq){r^2-\ell^2\over r^2+\ell^2}\right]. \label{ququlima} \eeq
($F$ and $f_r$ are positive when towards the center). Since this
configuration is obtained by inversion from one in which the large
 charge is at the center of the polygon, the above results also follow
 by applying the force-transformation formula
eq.(\ref{locurat}) to the results of the symmetric case.
Inversion about a point placed on a uniformly charged ring
 (limit of a polygon)
transforms it into a line with charge density
 $\r(x)=\r_0[1+(x/A)^2]^{-1}$, and the point charge at the center
 can be moved
to a point at an arbitrary distance in the symmetry plane of the wire.

\end{document}